\documentclass[usenatbib]{basi}
\usepackage[british]{babel}
\usepackage[varg]{txfonts}
\usepackage{rotating}
\usepackage{dcolumn}
\begin{document}
\title[Classification by Boosting Differences in Input Vectors:
An application to datasets from Astronomy]
{Classification by Boosting Differences in Input Vectors\thanks{By
            Ninan Sajeeth Philip, email: \texttt{nspp@iucaa.ernet.in}}}
\author[N. ~S. ~Philip et~al.]%
       {\small {N.~S.~Philip$^1$,
       A.~Mahabal$^{2}$, S. Abraham.$^1$, R. Williams${^2}$,
       S.G. Djorgovski$^{2,3}$, A. Drake${^2}$,
       C Donalek${^2}$ and M. Graham${^2}$}\\ 
       $^1$St. Thomas College, Kozhencheri, Kerala, India\\
       $^2$ Caltech, 1200 E California Bl., Pasadena, CA, 91125, USA\\
       $^3$ Distinguished Visiting Professor, King Abdulaziz University, Jeddah, Saudi Arabia}

\pubyear{2011}
\volume{00}
\pagerange{\pageref{firstpage}--\pageref{lastpage}}

\date{Received \today}

\maketitle
\label{firstpage}

\begin{abstract}
There are many occasions when one does not have complete information in order
to classify objects into different classes, and yet it is important to do the
best one can since other decisions depend on that. In astronomy, especially
time-domain astronomy, this situation is common when a transient is detected and
one wishes to determine what it is in order to decide if one must follow it. We
propose to use the Difference Boosting Neural Network (DBNN) which can boost
differences between feature vectors of different objects in order to
differentiate between them. We apply it to the publicly available data of the
Catalina Real-Time Transient Survey (CRTS) and present
preliminary results. We also describe another use with a stellar
spectral library to identify spectra based on a few features. The technique
itself is more general and can be applied to a varied class of problems.

%
%
\end{abstract}


\section{Introduction}\label{s:intro}

It is common to have fragmented and fragmentary evidence when the sources of
information are varied as well as unequal in strength. In time-domain astronomy,
just like in forensic science, one tries to piece together the information in
order to obtain a verdict. An object has been observed that was either not there
before, or was much fainter. The only extra information available is from
sporadic past observations of the same area in archival surveys that answer
questions like: `was this seen at such and such radio frequency?', `was it seen
in the Sloan Digital Sky Survey?', `what is the distance to the nearest
galaxy?' etc. Depending on the location of the object, and its nature, the
resulting information can be sparse and very different from source to source.
The Catalina Real-Time Transient Survey (CRTS\footnote{http://crts.caltech.edu})
publishes the transients it finds in real-time. The events, broadcast as
VOEvent packets, are ingested by Skyalert\footnote{http://www.skyalert.org}
\citep{2009ASPC..411..115W} where a
portfolio is gathered for each object by annotating the initial information
using programs that query individual surveys and archives to answer, as best as
possible, a pre-determined set of questions as mentioned earlier. If one thinks
of all such bits associated with an individual object as its feature,
the entire input data is just a vector of features (with many, often 50-80
percent, individual features for each vector missing). Making sense of this
dataset is not trivial, but classification is important 
(\citet{2012arXiv1111.3699M}, \citet{2011CIDU}).
Classifying the features into classes that are as unambiguous as
possible, and with minimal number  of  false detections needs a probabilistic
approach that can make full use of all available prior information.

The present work summarizes one possible method to handle such situations where
the sparseness of data is diverse and patchy whereby making it difficult
for existing machine learning techniques to handle them. It uses Bayesian
belief update rule that can be used to progressively update the belief in the
outcome based on plausible evidence. We apply the method to the CRTS
transients to generate quick classification probabilities which can be
revised as more data become available. We also show how it can be generalized
to be used with a stellar spectral library \citep{2004ApJS..152..251V} making it especially
useful in future when IFUs make available large number of spectra simultaneously.


\section{Catalina Real-Time Survey}\label{s:CRTS}

The Catalina Real-Time Transient Survey (\citet{2009ApJ...696..870D},
\citet{2011arXiv1102.5004D}, \citet{2011BASI...39..387M}) use data from the Catalina Sky Survey
(CSS\footnote{http://www.lpl.arizona.edu/css/}) for near-Earth objects and
potential planetary hazard asteroids (NEO/PHA), conducted by Edward Beshore,
Steve Larson, and their collaborators at the University of Arizona.  CRTS looks
for astrophysical transient and variable objects using real-time processing and
carries out characterization, and distribution of these events, as well as
follow-up observations of selected events.  

Optical transients (OTs) are detected as sources displaying significant
changes in brightness, as compared to the baseline comparison images and
catalogs with significantly fainter limiting magnitudes.  Data cover time
baselines from 10 min to several years.  The detected transients are published
electronically in real time, using a variety of mechanisms.  One of the methods
is a Skyalert stream where additional data on the
transients is gathered by harvesting archival datasets as well as information
on the proximity of the source in what is termed as passive follow-up. 

It is this comprehensive dataset that we make use of with DBNN in order to try to
predict the nature of the transient.  
The training set is based upon classification by human experts. Besides the extremely sparse matrix
visible to DBNN (described in the next section) the humans can use such aids as
the historic light curve of the transient which a human neural network can use
to trivially discriminate between SNe (single hump) versus a CV (multiple
brightenings over a few years) in many cases. Such features are being
incorporated into separate tools. Some derived characteristics from such aids
will be incorporated into DBNN in the future. The different techniques will
also be incorporated in to a fusion network.

\section{Difference Boosting Neural Network (DBNN)}\label{s:DBNN}

According to the Bayesian Theorem \citep{2011RvMP...83..943V}, it is possible
to start with an initial belief about the probability of occurrence of an event
even when there is no compelling evidence and later update this belief
periodically as new evidence is found. In the context of this paper, the
initial belief is called prior and the likelihood for an observation to cause
an event is called evidence. Then according to Bayesian theorem, popularly
called the Bayes rule, the updated belief, or the posterior, is the product of
the prior and the evidence normalized over all possibilities. Mathematically,
this may be written as: \begin{equation} P(A\vert B) = \frac{P(B\vert A)\times
P(A)}{\sum_i P(B\vert A_i)\times P(A_i)} \end{equation} where $P(A\vert B)$ is
the updated belief that the event A is caused by the occurrence of event B and
$P(B\vert A)$ is the likelihood that event B may cause A. P(A) is the prior,
the initial belief, which is independent of whether B is observed or not. One
may call it the probability that such an event may be found even when none of
the different types of
evidence is seen. The index $i$ is used to normalize all the possible events
including $A$ that could have produced $B$. 

The beauty of the Bayesian rule is that it allows sequential updating of the
belief or confidence in an outcome as more and more evidence arrives. Each time
an update is made, the computed new belief becomes the new prior for the next
update, thus essentially making the situation conducive for systems where one
has to deal with a diverse set of inputs and come up with plausible causes that
created them. The second advantage is that, all these estimates are based on
the statistical distribution of the likelihoods and hence the final evidence is
the probability and can be directly used as the confidence one may have in the
prediction of the Bayesian classifier.

We have used a Bayesian Classifier named \textit{Difference Boosting Neural
Network} \citep{2000cs........6001S} for this study. The DBNN distributes the
evidence in a feature space so that every feature is associated with its
likelihood as learned through a process called learning
\citep{2009NaBIC......4S, 2010PJBR........160S}. It also estimates the prior
for each evidence in individual cases that maximizes the prediction accuracy on
the data used for training. This data is referred to as training data. Assuming
that we have a large set of training data, it is possible to have a reasonable
estimate of the prior and the likelihood even for very complex cases. After the
training process, the estimated likelihoods and priors are saved 
for future use. 

Training is usually followed by a testing cycle in which the
classifier is tested with a fresh set of data that are similar but never used
in the training process. This is to test for adequate learning of the classifier,
in which case, the accuracy obtained on the training data and that on the test
data will be comparable. If that is not the case, one has to add the failed
examples also into the training data and update the likelihoods taking them
into consideration. This results in a \textit{continued learning process} that
Bayesian classifiers can very efficiently handle. The second advantage of this
progressive learning process is that the system asymptotically converges to a
global optimum as the learning progresses. This similarity of Bayesian rule to
human learning made Laplace comment that it is the \textit{mathematical
equivalent of common sense}.  

Preparing the data in a format required for use by the classifier is here referred to as
preprocessing. This is somewhat similar to preprocessing and data reduction
that are familiar to astronomers. The preprocessing ensures that the input
features are ordered in some fashion and an appropriate label is used to
represent the associated class of each example in the data. Usually the number
of input features may be fixed and the names of the classes all known. But the
new situation narrated above has no such restriction. One can have a dynamic
situation where new features are to be incorporated.  For managing this new situation, we have adopted a sequence coding method in which the presence or absence of a feature is marked by a 1 or 0 in the form of a chain of binary numbers. New entries are appended to the sequence when new evidence becomes available, allowing dynamical growth of the feature vector.

\begin{table}
\centering
\begin{tabular}{|c|l|}
\hline
Num. Code & Class \\
\hline
1 & Cataclysmic Variable\\
2 & Supernova\\
3 & other\\
5 & Blazar Outburst\\
6 & Active Galactic Nucleus Variability\\
7 & UVCeti Variable\\
8 & Asteroid\\
9 & Variable\\
10 & Mira Variable\\
11 & High Proper Motion Star\\
12 & Comet\\
16 & Nova\\
\hline
\end{tabular}
 \caption{The 12 different CRTS transient classes identified with numeric labels and common names. \label{f:class}}
\end{table}


The preprocessed data may have many missing entries depending on the list of
evidence available per feature vector.
The present scheme makes it possible for us to use
any additional information available about an entry to make more meaningful
prediction about its nature. The number of new entries
and the learning based on them dynamically vary as learning progress.
However, because all updating is done statistically, the subtle uncertainties
and errors in the features do not affect the decision
making process. The DBNN code is able to directly train and test on the
preprocessed data and we describe the application of the algorithm to
CRTS and stellar spectral studies in the following section.

\begin{table}
\begin{tabular}{|l|c|c|c|c|c|c|c|c|c|c|c|c|l|}
\hline
Real $\rightarrow$	& 1&	2 &	3 &5 & 	6 & 7 &8&	9& 10 & 11 & 12 & 16 & Total\\ \hline
 Predicted & & & & & & & & & & & & & \\
 1 &	 273&	 14 & 4 & 2	& 1& 4	& 3	& 3	& 6	& 1& 0& 0 &	 311 \\
 2 &	 15&    447 & 7	& 3	& 6& 2	& 4	& 5	& 1	& 4& 1& 2 &	 497 \\
 3 &	 1	& 1	& 47	& 0	& 1& 0	& 0	& 0	& 1	& 0& 0& 0 & 51 \\
 5 &	 0	& 0	& 0	& 68& 0	& 0& 0	& 0	& 0	& 0	& 0& 0& 68 \\
 6 &	 0	& 1	& 1	& 1	& 144 &	 1	& 1	& 3	& 0	& 1& 0& 0& 153 \\
 7 &	 0	& 0	& 0	& 0	& 0	& 33& 0	& 0	& 0	& 0	& 0& 0& 33 \\
 8 &	 0	& 0	& 0	& 0	& 0	& 0	& 5	& 0	& 0	& 0	& 0& 0& 5 \\
 9 &	 0	& 0	& 0	& 0	& 0	& 0	& 0	& 15& 0	& 0	& 0& 0& 15 \\
 10 &	 0	& 0	& 0	& 1	& 0	& 0	& 0	& 0	& 9& 0	& 0& 0& 10 \\
 11 	&2	& 0	& 1	& 0	& 0	& 0	& 0	& 0	& 0	& 51& 0& 0& 53 \\
 12 	&0	& 0	& 0	& 0	& 0	& 0	& 0	& 0	& 0	& 0	& 5& 0& 5 \\
 16 	& 0	& 0	& 0	& 0	& 0	& 0	& 0	& 0	& 0	& 0	& 0& 0& 0 \\ \hline
 Total	& 291	& 463	& 59	& 75&	 152	& 40	& 13&	 26&17	& 57& 6	& 2	& 1201\\ \hline
\end{tabular}\caption{The confusion matrix of the class labels: humans (Horizontal axis) against predicted (Vertical). \label{f:CRTSAgree}}
\end{table}

\section{Results}\label{s:Results}

The CRTS data that was used for this study has 12 distinct classification
labels enumerated in Table \ref{f:class} with their numerical class labels
followed by the common names. The input sequence used for this study had 39
input features (including detection magnitudes, colors from archival photometry,
distance to nearest star/galaxy etc.) 
of which 50 -- 80 percent were missing in some cases. None of
the cases had all 39 inputs. This forms only a subset of the data that human
experts usually use for discrimination.  We have not yet translated all the
available information that human experts use into machine recognizable format
(for example, archival light curves). This
is something we want to do in the near future.

Despite these limitations, the classifier agrees with human experts in
more than 90\% of the cases as the graduated training progresses and it learns about the 
variety in the evidence. This is shown in the confusion matrix Table
\ref{f:CRTSAgree}. The confusion matrix is a convenient representation showing
all relevant information such as how many objects in each class were correctly
identified and into which classes they were incorrectly labeled etc. It may
also be used to have a rough estimate of the number density of the different
classes by normalizing the totals in each class (last row) by the total number
of objects (last column of the last row).  

\begin{table}
\begin{tabular}{|c|l|c|c|c|}
\hline
Num. Code & Object Type & Total Objects & Completeness (\%) & Contamination (\%)\\
\hline
1 & Cataclysmic Variable & 291 & 93.8 & 12.2 \\
2 & Supernova & 463 & 96.5 & 10.1  \\
3 & Other &59 & 79.7 & 7.8  \\
5 & Blazar Outburst & 75 & 90.7 & 0.0 \\
6 & Active Galactic Nucleus Variability & 152 & 94.7 & 5.9 \\
7 & UVCeti Variable & 40 & 82.5 & 0.0 \\
8 & Asteroid & 13 & 38.5 & 0.0 \\
9 & Variable & 26 & 57.7 & 0.0 \\
10& Mira Variable & 17 & 52.9 & 10 \\
11& High Proper Motion Star & 57 & 89.5 & 3.7 \\
12& Comet & 6 & 83.3 & 0.0 \\
16& Nova & 2 & 0.0 & 0.0 \\
\hline
\end{tabular}
\caption{Completeness and contamination details of predictions by the DBNN classifier based on Table \ref{f:CRTSAgree} for the different type of objects in CRTS survey are shown.\label{f:complt}}
\end{table}

\begin{table}
\begin{tabular}{|c|l|c|c|c|}
\hline
Num. Code & Object Type & Total Objects & Completeness (\%) & Contamination (\%)\\
\hline
1 & Cataclysmic Variable & 61 & 63.9 & 35.0 \\
2 & Supernova & 74 & 74.3 & 40.9  \\
3 & Other & 14 & 0.0 & 100  \\
5 & Blazar Outburst & 9 & 55.6 & 16.7 \\
6 & Active Galactic Nucleus Variability & 26 & 76.9 & 48.7 \\
7 & UVCeti Variable & 4 & 25.0 & 0.0 \\
8 & Asteroid & 1 & 0.0 & 0.0 \\
9 & Variable & 3 & 0.0 & 0.0 \\
10& Mira Variable & 2 & 0.0 & 0.0 \\
11& High Proper Motion Star & 5 & 0.0 & 0.0 \\
12& Comet & 1 & 0.0 & 0.0 \\
16& Nova & 0 & 0.0 & 0.0 \\
\hline
\end{tabular}
\caption{Classification of transients from one of the test samples. Possible reasons
for relatively lower completeness are detailed in the text.\label{f:test}}
\end{table}

The performance of the classifier was quantified in terms of contamination and
completeness. The fraction of the total number of objects in a class that were
correctly recognized by the classifier is referred to as completeness and the
fraction that comes as contaminants into a class due to incorrect labelling by
the classifier is referred to as contamination. Both measures are indicative of
how reliable the classifier is. Table \ref{f:complt} shows the
completeness and contamination of the classifier predictions for the CRTS data
taking the verdict of the human expert as reference. 

\begin{figure}
\includegraphics[width=7 cm, height=5.3 cm]{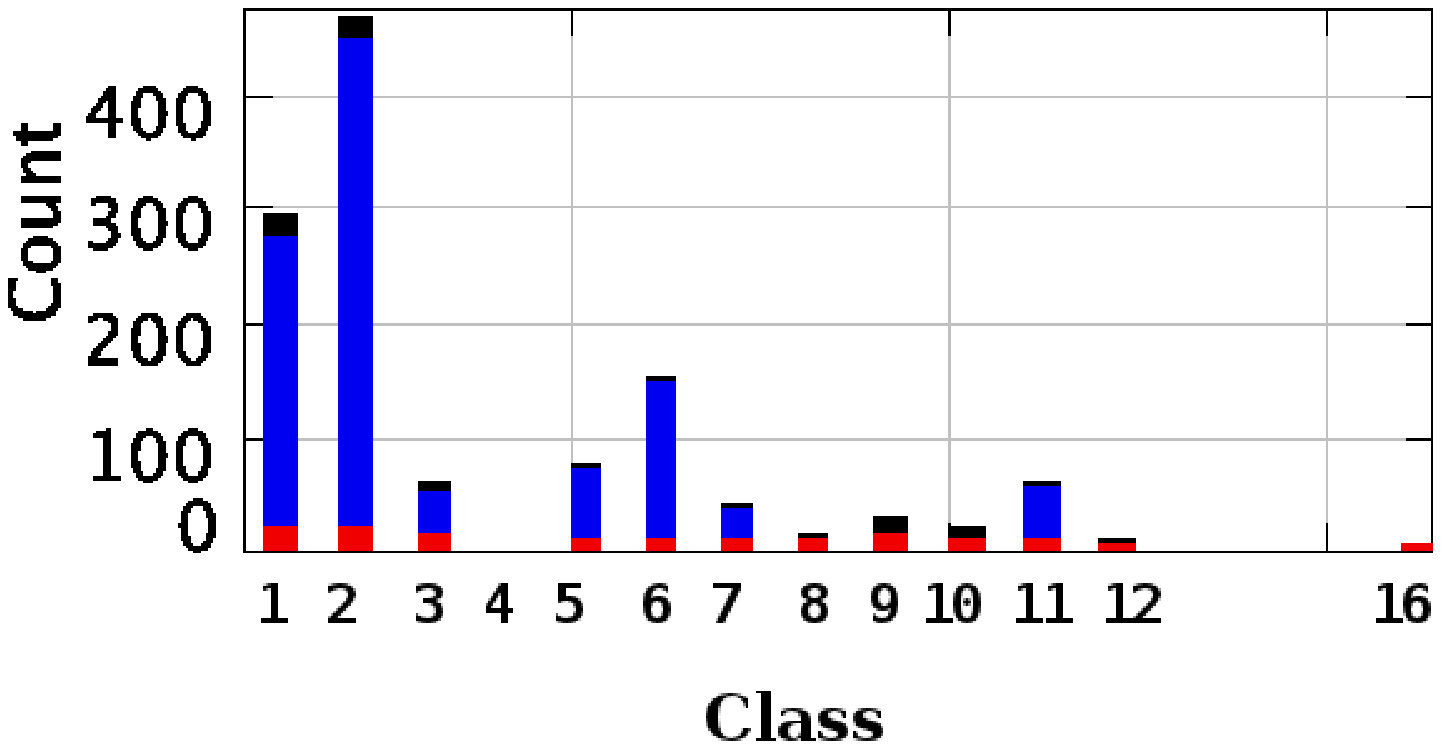}\includegraphics[width=7 cm, height=5.3 cm]{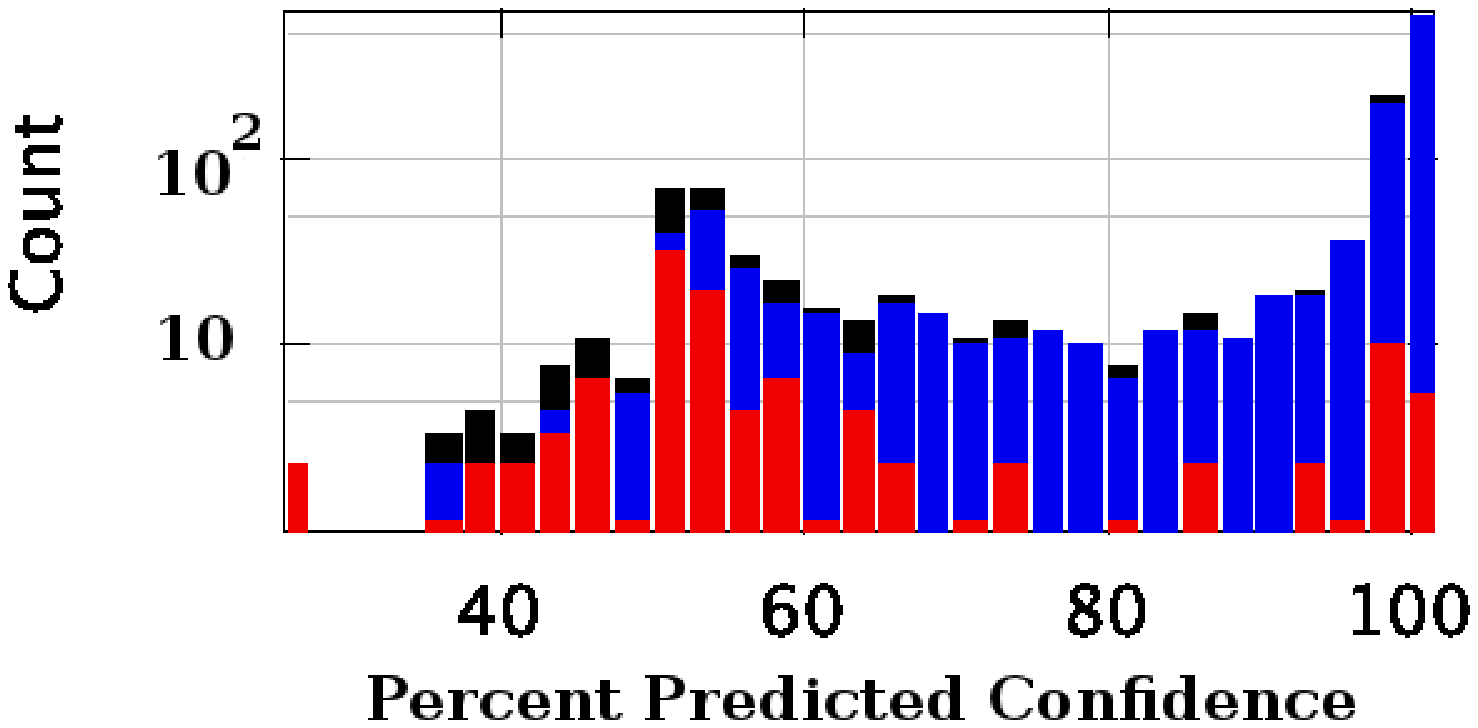} 
\caption{The plots show the performance of the classifier in predicting the
nature of the objects in the CRTS survey. In the plot to the left, black colour
is used to represent the total count of objects in each class while blue colour
is used to represent the number of correctly recognized objects and red colour
is used to indicate the number of contaminants in each class. The plot to the
right shows the same with percent Bayesian confidence and the counts (in log scale to improve clarity). }\label{fig:Conf}
\end{figure}

As explained earlier, the classifier evaluates the Bayesian posterior
probability that can be considered synonymous to the true confidence the
classifier predicts. A graphical representation of the
completeness and confidence measures gives a more intuitive picture of the
classifier performance (Fig. \ref{fig:Conf}).  It may be noted
that for most of the failed cases, the confidence was low indicating that they
are not as reliable as the rest.
It is advisable to use this as a
guide to determine how safe it is to rely on low confidence predictions and to
draw a cut-off line which states that anything below is unreliable. 
Quantification of this is in progress as more data come in.

We have also carried out tests with training the method on a smaller subset and testing on feature vectors never seen before. Due to the sheer variety in features, and the fact that a large majority of them are missing, the performance worsens (Table \ref{f:test}). One other important reason for this is that a few of the classes, like asteroid, comet and high proper motion stars (HPM) are not static and hence the archives do not have any information about them (at the discovery location). To incorporate that, additional features will have to be introduced. Even for many of the other classes there are not enough examples as yet. Including archival light curves will help greatly improve the SNe/CV dichotomy. All that is part of on going research, and that should be kept in mind when judging the current results.

As another example, the classifier was used on stellar spectra to classify them
into 98 different classes as given in the Indo-US Stellar
library\footnote{http://www.noao.edu/cflib/} \citep{2004ApJS..152..251V}. The
input features used were the major absorption lines in the spectra and the
maximum flux values from four regions of the continuum in a window of {200\AA}
centered around wavelengths  3700, 4500, 6300, {8500\AA}, respectively. These
spectra have a coverage from {3460\AA}  to {9464\AA} with a few that have missing
bands in between. It was taken with a 0.9m Coud$e^{'}$ telescope at Kitt Peak
National Observatory in five different grating settings. 

The sequence length of the extracted features used by us for this study had  32
most distinctive features in the spectra quantified by their equivalent widths
and the said four flux values. We used 958 examples for training the classifier
and 1104 examples for testing the learning. It was found that 88\% of the
classifications were in agreement with the classifications given in the
catalog. The purpose of this work was only to demonstrate that the method
might find some application in spectral analysis, especially for chemical
abundance measures in the stellar atmosphere at a larger scale than what is
possible otherwise. 

\section{Conclusions}\label{s:Conclusions}
We describe a new method to make use of diverse and sparse information from
various sources to classify astronomical data. One practical application we
found is in the case of transients where alerts need to be sent to astronomers
to carry out follow up observations whenever an object that is likely to be of
interest to them is detected. However, it is possible to extend the method to
other applications such as spectroscopic classification where it is difficult
to predefine important absorption/emission lines and we want to cluster them in
a high dimensional space. 


\section*{Acknowledgements}

The work at Caltech has been supported in part by the NSF grants AST-0407448, CNS-0540369, AST-0834235, AST-0909182 and IIS-1118041; the NASA grant 08-AISR08-0085; and by the Ajax and Fishbein Family Foundations.
The first author wishes to thank Prof. Ranjan Gupta for useful discussions and acknowledges the use of Indo-US Library of Coud$e^{'}$ Feed Stellar Spectra.
\end{document}